\documentclass[aip,amsmath,amssymb,reprint]{revtex4-1}

\usepackage{graphicx}% Include figure files
\usepackage{xcolor}
\usepackage{dcolumn}% Align table columns on decimal point
\usepackage{bm}% bold math
\usepackage[utf8]{inputenc}
\usepackage[T1]{fontenc}
\usepackage{mathptmx}
\usepackage{etoolbox}
\usepackage{hyperref}
\usepackage{overpic}

\renewcommand{\d}{{\rm d}}
\newcommand{\half}{{\frac{1}{2}}}

\newcommand{\ignore}[1]{}

\newcommand{\Btree}{{\mathbf{B}_\wedge}}
\newcommand{\Btri}{{\mathbf{B}_\Delta}}

\makeatletter
\def\@email#1#2{%
 \endgroup
 \patchcmd{\titleblock@produce}
  {\frontmatter@RRAPformat}
  {\frontmatter@RRAPformat{\produce@RRAP{*#1\href{mailto:#2}{#2}}}\frontmatter@RRAPformat}
  {}{}
}%
\makeatother
\begin{document}

\preprint{AIP/123-QED}

\title[Vascular networks: towards model calibration]{
Bifurcations in adaptive vascular networks: towards model calibration
}
\author{Konstantin Klemm}
\email{kklemm@posteo.net}
\affiliation{Instituto de F\'isica Interdisciplinar y Sistemas Complejos (IFISC, CSCIC-UIB), Campus Universitat de les Illes Balears,
E-07122 Palma de Mallorca, Spain}

\author{Erik A. Martens}%
\email{erik.martens@math.lth.se}
\affiliation{ Centre for Mathematical Science, Lund University, Sölvegatan 18B, 22100, Lund, Sweden}%

\date{\today}% It is always \today, today,

\begin{abstract}
Transport networks are crucial for the functioning of natural and technological systems. We study a mathematical model of vascular network adaptation, where the network structure dynamically adjusts to changes in blood flow and pressure. The model is based on local feedback mechanisms that occur on different time scales in the mammalian vasculature. The cost exponent $\gamma$ tunes the vessel growth in the adaptation rule, and we test the hypothesis that the cost exponent is $\gamma= 1/2$ for vascular systems [Hu and Cai, Phys. Rev. Lett., Vol. 111(13) (2013)~\cite{HuCai2013}]. We first perform a bifurcation analysis for a simple triangular network motif with fluctuating demand, and then conduct numerical simulations on network topologies extracted from perivascular networks of rodent brains. We compare the model predictions with experimental data and find that $\gamma$ is closer to 1 than to 1/2 for the model to be consistent with the data. Our study thus aims at addressing two questions: (i) Is a specific measured flow network consistent in terms of physical reality? (ii) Is the adaptive dynamic model consistent with measured network data? 
We conclude that the model can capture some aspects of vascular network formation and adaptation, but also suggest some limitations and directions for future research. Our findings contribute to a general understanding of the dynamics in adaptive transport networks, which is essential for studying mammalian vasculature and developing self-organizing piping systems.
\end{abstract}

\maketitle

\section{Introduction}
% \begin{enumerate}
%  \item General introduction:
%  \item [-] Transport and flow networks. 
%  \item [-] Optimal network structure and self-organizing adaptive (flow) networks.
%  \item Problems/challenges 
%  \item[-] I: Consistency of measured flow networks?
%  \item [-] II: Consistency of adaptive dynamic model with measured network data.
% \end{enumerate}

Transport and flow networks are ubiquitous in nature and engineering, ranging from blood vessels in  mammals, ~\cite{Smith2006,Postnov2016}, glymphatic fluid networks in the brain~\cite{kelley2022glymphatic}, sap flow in trees, plant roots, and leaf venation~\cite{jensen2016sap} to roads, pipelines and fluid microcircuits~\cite{martinez2023fluidic}. These networks have to efficiently distribute resources from sources to sinks under varying demands and constraints. One way to achieve this efficiency is by self-organizing adaptive networks, where the network structure dynamically adjusts to the local flow conditions~\cite{}. Such networks can exhibit complex topologies that include cycles and loops, which may enhance robustness and flexibility, as shown by previous studies using optimization methods~\cite{Katifori2010,Dodds2010,Farr2014,ronellenfitsch2019phenotypes}.

However, there are two main challenges in understanding and modeling self-organizing adaptive networks. The first one is to obtain reliable measurements of the network structure and dynamics in real systems, such as the vascular system of animals or plants~\cite{Smith2006,katifori2012quantifying,kirst2020mapping}. The second one is to develop and validate mathematical models that can reproduce the observed network features and predict their behavior under different scenarios.

In this article, we address these challenges for a specific class of self-organizing adaptive networks: vascular network systems. We focus on the calibration of a model that describes how the network adapts to changes in blood flow and pressure, based on local feedback mechanisms~\cite{Jacobsen2009}. Such mechanisms are known to occur --- in general --- on a range of time scales in the mammalian vasculature, ranging from seconds (myogenic responses induced by smooth muscle cells encompassing arterioles) to months (remodeling of blood vessels)~\cite{Jacobsen2008,Jacobsen2009}. In particular, we build on a mathematical model of vessel adaptation originally proposed by Hu and Cai~\cite{HuCai2013}. This model produces results from our previous studies as a limiting case~\cite{MartensKlemm2017,martens2019cyclic}. 
The network structure emerging from this adaptive model depends on the spatial distribution of sources and sinks, and on the amplitude and frequency of flow fluctuations. The model can capture certain aspects of vascular network formation and adaptation, such as the emergence of cycles, the scaling of vessel diameters, and the balance of pressure fluctuations~\cite{HuCai2013,MartensKlemm2017,martens2019cyclic}. 
An important model parameter is the exponent $\gamma$ in Eq.~\eqref{eq:adaptationrule} which tunes between a concave versus convex vessel growth in the adaptation rule. Hu and Cai argued~\cite{HuCai2013} that this exponent should be $\gamma = 1/2$ for vascular systems. To test this hypothesis, we first carried out a bifurcation analysis for a simple triangular network motif, composed of one constant source and two sinks subject to fluctuating demand, in dependence on fluctuation amplitude and the cost exponent $\gamma$. We then conducted numerical simulations on the model on network topologies extracted from perivascular networks of rodent brains~\cite{Blinder2010} to study the emergent network structure as vessels appeared or vanished while varying the cost exponent $\gamma$. 
Thereby we aim at addressing two interlinked questions: 
\begin{itemize}
 \item [(i)] Is a specific measured flow network consistent in terms of physical reality? 
\item [(ii)] Is the adaptive dynamic model consistent with measured network data?
\end{itemize}
We compared model predictions with experimental data from the perivascular network of rodent brains~\cite{Blinder2010}. 
However, our comparison of model predictions with experimental data suggest that --- in contrast to Hu and Cai's study~\cite{HuCai2013}--- $\gamma$ is closer to 1 than to $1/2$ when chosen such that the model is able to sustain the topological structure found in real world networks.

The article is structured as follows. Section~II explains the mathematical model of adaptation and briefly discusses the nature of the vascular network data used for comparing model predictions. Section~III presents a detailed bifurcation analysis for the triangular motif; and discusses the numerical simulation on the perivascular rodent brain networks, and Section~IV concludes our study with a discussion.

\section{Model, network data, and methods}

\subsection{The model by Hu and Cai}
Let $V$ denote a set of nodes of a network with $N = |V| < \infty$ and $A \subseteq N \times N$ the set of edges. The edges are bidirectional, so $(i, j) \in A$ implies $(j, i) \in A$. Each node is assigned a pressure $p_i$. The edge flow is $Q_{ij} > 0$ from node $i$ to $j$. We assume that the network is resistive and linear, i.e., Ohmian with $Q_{ij} = C_{ij}(p_i - p_j)$, where an edge carries the property of a conductance between nodes $i$ and $j$ with $C_{ij} = C_{ji} > 0$ only if $(i,j)\in A$. Furthermore, we assign a length $L_{ij}$ to each vessel segment between node $i$ and node $j$. 

At each node $i$, Kirchhoff's law (mass balance) demands that
\begin{align}\label{eq:massbalance}
h_i = \sum_{j \in V} Q_{ij} = \sum_{j \in V} C_{ij} (p_i - p_j)
\end{align}
with the flow $Q_{ij} = C_{ij} (p_i - p_j)$ and the constraint
$\sum_{i \in V} h_i =0.$

Hu and Cai~\cite{HuCai2013} proposed an adaptation model for biological transport networks
\begin{align}\label{eq:adaptationrule}
\frac{\d }{\d t} \tilde C_{ij} = c \left( \frac{Q_{ij}^2}{\tilde C^{\gamma+1}} - \tau_e^2 \right) \tilde C_{ij}
\end{align}
By appropriate choice of units of time and conductance, we obtain values of the constants $c=\tau_e=1$. Then, since  $Q_{ij} = C_{ij} (p_i - p_j)$, we have
\begin{align}
\frac{\d }{\d t} \tilde C_{ij} = \frac{C_{ij}^2 (p_i-p_j)^2}{\tilde C^{\gamma}} - \tilde C_{ij},
\end{align}
where assume that the cost exponent is positive, $\gamma>0$.
Finally, substituting $\tilde C_{ij} = C_{ij}L_{ij}$, we arrive at
\begin{align}\label{eq:adaptation}
\frac{\d }{\d t} C_{ij} = C_{ij}^{2-\gamma}(p_i-p_j)^2 L_{ij}^{-1-\gamma} - C_{ij}.
\end{align}

\subsection{Sink fluctuations}

We consider systems with a single source node $r \in V$ having $h_r = 1$.
We denote the set of sink nodes with $S \subset V$ where $r \notin S$. The net flow at sink nodes is assumed to fluctuate in general. This is implemented by choosing one sink $s \in S$ and assigning $s$ a larger net flow (in absolute value) than the the other sinks as
\begin{align}
  h_i = \begin{cases}
                       1 & \text{ if } i=r \\ 
  -\dfrac{1-\alpha}{|S|}-\alpha & \text{ if } i=s \\
  -\dfrac{1-\alpha}{|S|}        & \text{ if } i \in S\setminus\{s\}\\
  0                      & \text{ otherwise. }
\end{cases} 
\end{align}
The parameter $\alpha$ determines the amplitude of fluctuations. For $\alpha=0$, fluctuations vanish with each sink $i \in S$ having constant outflow $h_i=-1/|S|$. The case of $\alpha =1$ is a single moving sink since all sinks except high-flow sink $s$ have have zero net flow. The high-flow sink $s$ is redrawn uniformly at random after time intervals shorter than the time scale of the conductances adapting by \eqref{eq:adaptationrule}, see Section~\ref{sec:1source2sinks} for details.

\subsection{Pial network of neocortex in rodents} \label{sec:networkdata}

As an empirical testbed for the model, we employ a vascular network from the neocortex in a rodent \cite{Blinder2010}. The network with identifier 012208 has 855 connections (vessel segments) connecting 826 nodes, 416 of which are sink nodes. A node $i$ is a sink node, $i \in S$, if and only if $i$ has degree 1 or degree 2 meaning that fewer than three vessel segments meet at $i$. In addition to the list of node pairs forming connections, the data contain the coordinates $(x_i,y_i)$ in a 2-dimensional Euclidian space for each node $i$. The length of a vessel segment between nodes $i$ and $j$ is given by $L_{23}=\sqrt{(x_i-x_j)^2+(y_i-y_j)^2}$ being the Euclidian distance between the nodes. Here we assume that vessel segments have neglibible curvature thus extending along a straight line.

\section{Results}

\subsection{A system with 1 source and 2 sinks} \label{sec:1source2sinks}
% \emph{Calculations are replicated in Mathematica notebook CalibratingVascularNetworks\_FixedPointsn.nb.}

We consider a triangular motif with one source $h_1=1$ and two fluctuating sinks $h_2$ and $h_3$. 
We assume that the drive $h_{2,3}(t)$ is rapid with a well defined time scale, i.e., $T=T(h_{2,3})\ll 1$.
The sources then obey $\langle h_2 \rangle-\langle h_3 \rangle  \rightarrow 0$ and no net pumping occurs between nodes $k=2$ and $k=3$. 

Accordingly, we seek solutions, $\langle C_{kl}\rangle$, averaged over rapid fluctuations with characteristic time scale $T$. We consider that $C_{ij}$ changes on a slow time scale, $C_{ij} \rightarrow \langle C_{ij}\rangle$ as $T \rightarrow 0$. We therefore may from now on use $\langle C_{kl}\rangle$ and $C_{kl}$ interchangeably; to simplify notation we henceforth omit $\langle\cdot\rangle$ around the conductances. The symmetry of the triangular network motif and the rapid driving implies that $\langle C_{12}\rangle = \langle C_{13} \rangle$. In this limit, the dynamics of the conductances is constrained to a two dimensional subspace on which the dynamics is given by
\begin{subequations}
\begin{align}
\label{eq:triangular_motif_goveqn_symmetric_a}
 \frac{\d }{\d t} C_{12} &=  C_{12}(C_{12}^{1-\gamma}\langle(p_1-p_2)^2\rangle L_{12}^{-1-\gamma} - 1),\\
 \label{eq:triangular_motif_goveqn_symmetric_b}
 \frac{\d }{\d t} C_{23} &= C_{23}(C_{23}^{1-\gamma}\langle (p_2-p_3)^2 \rangle L_{23}^{-1-\gamma} - 1).\
\end{align} 
\end{subequations}
The conductances are related via mass balance \eqref{eq:massbalance}, i.e.,
\begin{subequations}
\begin{align}
\label{eq:triangular_motif_massbalance_a}
 1&= C_{12}(p_1-p_2)+C_{13}(p_1-p_3),\\
\label{eq:triangular_motif_massbalance_b}
 h_2&= C_{12}(p_2-p_1)+C_{23}(p_2-p_3),\\
\label{eq:triangular_motif_massbalance_c} 
 h_3&= C_{13}(p_3-p_1)+C_{23}(p_3-p_2),\
\end{align}
\end{subequations}
where we may set the reference pressure $p_1=0$ without loss of generality.
The symmetry $C_{12}=C_{23}$ simplifes the mass balance  \eqref{eq:triangular_motif_massbalance_a}-\eqref{eq:triangular_motif_massbalance_c} which, using $\xi:= \dfrac{C_{23}}{C_{12}+2C_{23}}$,  yields the following relations,
\begin{subequations}
\begin{align}
 p_2-p_1 &= \frac{1}{C_{12}}\left(h_2-(h_2-h_3)\xi\right),\label{eq:triangular_motif_massbalance_2_a}\\
 p_2-p_3 &= \frac{h_2-h_3}{C_{12}+2C_{23}} \label{eq:triangular_motif_massbalance_2_b},\
\end{align}
\end{subequations}
which allow us to eliminate the pressure differences in \eqref{eq:triangular_motif_goveqn_symmetric_a} and \eqref{eq:triangular_motif_goveqn_symmetric_b},
\begin{subequations}
\begin{align}
 \frac{\d }{\d t} C_{12} &=  C_{12}\left(C_{12}^{-1-\gamma}L_{12}^{-1-\gamma}\nonumber
  \left(\langle h_2^2\rangle -2\langle (h_2 (h_2-h_3)\rangle\xi \right.\right. \\
  &\quad \quad \quad \quad + 
  \left.\left. \langle(h_2-h_3)^2\rangle\xi^2\right) - 1\right),\\
 \frac{\d }{\d t} C_{23} &= C_{23}\left(C_{23}^{1-\gamma}L_{23}^{-1-\gamma} \frac{\langle (h_2-h_3)^2 \rangle}{(C_{12}+2C_{23})^2}  - 1\right).\  
\end{align}
\end{subequations}
Fluctuations depend on their amplitude, $\alpha$, and are given in the Appendix~\ref{app:fluctuations} in  Eqs.~\eqref{eq:fluctuations_periodic} and ~\eqref{eq:fluctuations_stochastic}.

To calculate the Jacobian, we introduce the short notation  $x:=C_{12}$ and $y:=C_{23}$. We then have
\begin{subequations}
\begin{align}
 \frac{\d }{\d t} y &= f(x,y),\quad
 \frac{\d }{\d t} x = g(x,y).\  
\end{align}
\end{subequations}
where
\begin{align*}\nonumber
  f(x,y) &= a x^{-\gamma}\left[\langle h_2^2\rangle -2\langle (h_2 (h_2-h_3)\rangle\xi + \langle(h_2-h_3)^2\rangle\xi^2\right]
   - x\\
  g(x,y)&= b y^{-\gamma}\langle (h_2-h_3)^2 \rangle\xi^2  - y\
\end{align*}
where $\xi= \frac{y}{x+2y}$, $a:=L_{12}^{-1-\gamma}$ and $b:=L_{23}^{-1-\gamma} $.
The Jacobian has the following entries:
\begin{subequations}
\begin{align}\label{eq:Jacobian_a}\nonumber
  f_{x}&= -\gamma a x^{-1-\gamma}\left[\langle h_2^2\rangle -2\langle (h_2 (h_2-h_3)\rangle\xi + \langle(h_2-h_3)^2\rangle\xi^2\right]\\
  &+2a x^{-\gamma}\left[\langle (h_2 (h_2-h_3)\rangle\xi_x - \langle(h_2-h_3)^2\rangle\xi\xi_x\right] - 1\\
   f_y&=-2a x^{-\gamma}\left[ \langle (h_2 (h_2-h_3)\rangle\xi_y + \langle(h_2-h_3)^2\rangle\xi\xi_y\right]\\
   g_x&=2 b y^{-\gamma}\langle (h_2-h_3)^2 \rangle\xi\xi_x  \\ 
   \label{eq:Jacobian_d}
   g_y&=-\gamma b y^{-1-\gamma}\langle (h_2-h_3)^2 \rangle\xi^2  
   +2 b y^{-\gamma}\langle (h_2-h_3)^2 \rangle\xi\xi_y  
   - 1\ 
\end{align}
\end{subequations}
where we note that 
$\xi_x = -y(x+2y)^{-2}$ and $\xi_y = (x+2y)^{-1}-2y(x+2y)^{-1}$.

\subsubsection{Stationary Solutions}
We denote solution branches by $\mathbf{B}=(C_{12},C_{23})$ where we generally assume $C_{12},C_{23} \geq 0$. 
The following stationary solutions are possible. 
We always require $C_{12}>0$; otherwise, no net transfer of mass is possible along the network, contradicting mass balance.  Solutions can either be \emph{'tree-like'} if $C_{23}=0$, i.e.,  $\Btree=(C_{12},0)$; or \emph{'triangular'} if $C_{23}>0$, i.e., $\Btri=(C_{12},C_{23})$. 
We consider only $\gamma>0$, and Eqs.~\eqref{eq:triangular_motif_goveqn_symmetric_a} and \eqref{eq:triangular_motif_goveqn_symmetric_b} inform us that tree-like solutions exist only if $0< \gamma \leq 2$ (note that $x^0\to 1 $ as $x\to 0$); if, on the other hand, $\gamma > 2$ the right hand sides of \eqref{eq:triangular_motif_goveqn_symmetric_b} are indeterminate $C_{23}=0$ (or infinite if $C_{23}\to 0^+$). Thus, we consider tree like solutions for $0< \gamma\leq 2$.

\subsubsection{Tree-like solution.} Applying the tree-like solution to \eqref{eq:triangular_motif_massbalance_2_a} we have $p_2-p_1 = h_2/C_{12}$. 
Imposing further a stationary nontrivial solution for \eqref{eq:triangular_motif_goveqn_symmetric_a} yields $1=C_{12}^{1-\gamma} \langle(p_1-p_2)^2 \rangle = C_{12}^{-1-\gamma} \langle h_2^2 \rangle$, so that $\langle h_2^2\rangle=x^{1+\gamma}=C_{12}^{1+\gamma}$ or $x=C_{12}=\langle h_2^2\rangle^{\frac{1}{1+\gamma}}$.
The tree-like branch is therefore given by $\mathbf{B}_\wedge=(C_{12},C_{23})=(\langle h_2^2\rangle^{\frac{1}{1+\gamma}},0)$.

We evaluate the Jacobian \eqref{eq:Jacobian_a}-\eqref{eq:Jacobian_d} for the tree-like branch. Using $y=C_{23}=0$ we have $\xi=0$, $\xi_x = 0$ and $\xi_y=1/x$. 
For $f_x$ all terms cancel that are proportional to both powers of $x$ and $\xi$, $\xi^2$; thus $f_x^\Btree=-1-\gamma$. 
\\
Similarly, for $f_y$ we only retain the term proportional to $\xi_y=1/x$; we use that $\xi_y \cdot x^{-\gamma}= (\langle h_2^2\rangle )^{-1}$   and obtain $f_y^\Btree=- 2a - a\langle h_2 h_3 \rangle / \langle h_2^2 \rangle$.
\\
More care is needed to evaluate $g_x$ and $g_y $, where products and powers of $y$ appear together with terms proportional to $y$, such as  $\xi,\xi_x,\xi_y \propto y$, thus resulting in terms of the form $y^p$ with some exponent $p$. If $p>0$, associated factors become zero as $y\to 0$; but if $p<0$ the associated term is indeterminate. 
Consider first $g_x\propto y^{2-\gamma}$. Since $y=0$ on the tree-like branch, we have $g_x^\Btree=0$ for $\gamma\leq 2$. It follows that the associated Jacobian is an upper triangular matrix.
\\
Next consider $g_y$, for which the first term is $y^{-1-\gamma}\xi^2\propto y^{1-\gamma}$, and the second is $y^{-\gamma}\xi\xi_y\propto y^{1-\gamma } + y^{2-\gamma}$.  Provided that $\gamma<1$, $g_y$ is determinate for $y=0$, in particular, all terms go to 0 as $y\to 0$, and we obtain $g_y^\Btree=- 1$.
% - y^{1-\gamma} (\gamma-2)
% \langle{h_2-h_3)^2\rangle / \langle h_2^2 \rangle ^{2/(1+\gamma)} }$ which is determinate provided that $\gamma<1$, so that $g_y^\Btree=- 1$. 
For $\gamma=1$, we use that the term with $y^{1-\gamma}=y^0\to 1$ as $y\to0^+$ (all other terms containing $y$ are 0), and we get $g_y^\Btree=- 1 +
\langle{h_2-h_3)^2\rangle / \langle h_2^2 \rangle ^{2/(1+\gamma)} }$.
Finally, considering $1\leq \gamma<2$ the expression $y^{1-\gamma}\rightarrow + \infty$ (as long as $\gamma<2$) as $y \rightarrow 0^+$ so that we have $g_y\rightarrow+\infty$. 

Since the Jacobian is triangular, we can just identify the eigenvalues as $\lambda_1^\Btree = f_x|_{\Btree}$ and $\lambda_2^\Btree = g_y|_{\Btree}$.
% Thus, the tree-like solutions are always stable. Unlike the triangular solutions, the tree-like solution is unique.
Thus, the eigenvalues for the tree like branch are
% \textcolor{red}{Valid for $0<\gamma<2$ or $0<\gamma\leq2$? Consider limit in Jacobian for $\gamma=2$.}
\begin{subequations}
\begin{align}
\lambda_1^\Btree&= - 1 - a\gamma \\
\lambda_2^\Btree&= 
    \begin{cases}
        -1&0<\gamma < 1,\\
        - 1 + \langle{h_2-h_3)^2\rangle / \langle h_2^2 \rangle ^{2/(1+\gamma)} } &\gamma=1,\\
        +\infty & 1<\gamma \leq 2.\
    \end{cases}
\end{align}
\end{subequations}

Note that letting $\gamma = 1$ and  using \eqref{eq:fluctuations_stochastic} we have 
% $\lambda_2^\Btree = \frac{6 a^2-1}{2 a^2+1}$. 
$\lambda_2^\Btree = \frac{3 \alpha^2-1}{\alpha^2+1}$. 
This determines a critical strength of fluctuations for 
% $a_c=1/\sqrt{6}$
$\alpha_c=1/\sqrt{3}$, where the tree-like branch loses stability in a transcritical bifurcation. Indeed this is the same result we obtained in a previous study~\cite{MartensKlemm2017}.

To sum up, the tree-like branch is always stable for $0<\gamma<1$; for $\gamma=1$ only if $\alpha<\alpha_c$; and for $1<\gamma\leq 2$ it is ``hyper-repellent''.

\subsubsection{Triangular solutions}

Let us turn our attention to triangular solutions where the cross-connection between the two sinks has positive conductance $C_{23} >0 $. We study the case with symmetry under exchange of the sink nodes where $C_{12} = C_{13}$ and $L_{12}=L_{13}=1$. We set the pressure at the source to zero and denote by $p_l$ the pressure at the sink with lower outflow, by $p_h$ the pressure at the sink with higher outflow. According to mass balance~\eqref{eq:massbalance}, the pressures satisfy
\begin{align}
  -\frac{1-\alpha}{2} & =   C_{12}\, p_l +C_{23}(p_l-p_h), \\
  -\frac{1+\alpha}{2} & =   C_{12}\, p_h +C_{23}(p_h-p_l),
\end{align}
which results in
\begin{align}
  p_l + p_h &= -\frac{1}{C_{12}}~,\\
  \qquad p_l - p_h &= \frac{\alpha}{C_{12} + 2 C_{23}}~.
\end{align}
This leads to the quadratic pressure differences
\begin{align}\label{eq:pl_ph}
  (p_l-p_h)^2 = \frac{\alpha^2}{(C_{12} + 2 C_{23})^2} =: X
\end{align}
and
\begin{align}\label{eq:p2_p1}
    \begin{split}
    \langle (p_2 - p_1)^2 \rangle = \frac{1}{2} (p_l^2+p_h^2) 
    &= \frac{1}{4} [(p_l + p_h)^2 + (p_l - p_h)^2]\\
    &= \frac{1}{4} \left(\frac{1}{C_{12}^2} + X\right)
    \end{split}
\end{align}
using $p_1=0$. 
% % 
We now look for stationary solutions of the conductances' dynamics \eqref{eq:adaptation}. Using~\eqref{eq:pl_ph}, the fixed point condition ${\dot C}_{23}=0$ yields
\begin{align}\label{eq:stationary23}
    \begin{split}
    0 & = C_{23}^{2-\gamma} (p_l-p_h)^2 L_{23}^{-1-\gamma}  - C_{23}\\
        & = C_{23}^{2-\gamma} X L_{23}^{-1-\gamma}  - C_{23}.
    \end{split}
\end{align}
With the assumption $C_{23} \neq 0$, \eqref{eq:stationary23} implies
\begin{align}\label{eq:XfromC23}
X = C_{23}^{\gamma-1} L^{\gamma+1}~.
\end{align}

Setting ${\dot C}_{12}=0$ gives
\begin{align}
    \begin{split}
  0 & = C_{12}^{2-\gamma} \langle (p_2-p_1)^2 \rangle - C_{12} \\
    & = \frac{1}{4} C_{12}^{2-\gamma} \left(\frac{1}{C_{12}^2} + X\right) - C_12 \\
    & = \frac{1}{4} C_{12}^{2-\gamma} \left(\frac{1}{C_{12}^2} + C_{23}^{\gamma-1} L^{\gamma+1}\right) - C_{12}  \\ 
    & = \frac{1}{4} C_{12}^{-\gamma} + \frac{1}{4} C_{12}^{2-\gamma} C_{23}^{\gamma-1} L_{23}^{\gamma+1} - C_{12} \label{eq:stationary12}
    \end{split}
\end{align}
with the help of \eqref{eq:p2_p1} and \eqref{eq:XfromC23}.

\paragraph{Case $\gamma = 1$.}
In the special of $\gamma = 1$, \eqref{eq:XfromC23} gives
\begin{align}
(C_12 + 2C_{23})^2 = \frac{\alpha^2}{L_23^2}
\end{align}
after inserting $X$ as defined in \eqref{eq:pl_ph}. Since both $\alpha$ and $L_{23}$ are non-negative, we may take square root and restrict to the positive branch to obtain
\begin{align}\label{eq:C23forgamma1}
C_{23} = \frac{1}{2} \left(\frac{\alpha}{L_{23}}-C_{12}\right)~.
\end{align}

The value of $C_12$ follows from setting $\gamma=1$ and multipliying by $4C_{12}^{-1}$ in \eqref{eq:stationary12}. We obtain
\begin{align}\label{eq:C12forgamma1}
C_{12} =\sqrt{(4-L_{23}^2)^{-1}}
\end{align}
for general $L_{23}$; and $C_{12}=1 / \sqrt{3}$ for the special case of the equilateral triangle system with $L_23 =1$. Inserting \eqref{eq:C12forgamma1} into \eqref{eq:C23forgamma1}, we arrive at
\begin{align}\label{eq:C23forgamma1final}
C_{23} = \frac{1}{2} \left(\frac{\alpha}{L_{23}}-\sqrt{(4-L_{23}^2)^{-1}}\right)
\end{align}
evaluating to
\begin{align}\label{eq:C23forgamma1finalequi}
C_{23} = \frac{1}{2} \left(\alpha-\sqrt{\frac{1}{3}}\right)
\end{align}
in the equilateral case of $L_{23}=1$. The solution \eqref{eq:C23forgamma1final} determines a critical parameter value $\alpha_c$ where the solution changes from negative to positive values (transcritical bifurcation). We have 
\begin{align}
\alpha_c = \sqrt{(4L_{23}^{-2} -1)^{-1}}
\end{align}
for general $L_{23}$; and $\alpha_c = \sqrt{1/3}$ in the equilateral case where $L_{23} =1$, thus reproducing our previous result~\cite{MartensKlemm2017,martens2019cyclic}.

\begin{figure}
\includegraphics[width=\columnwidth]{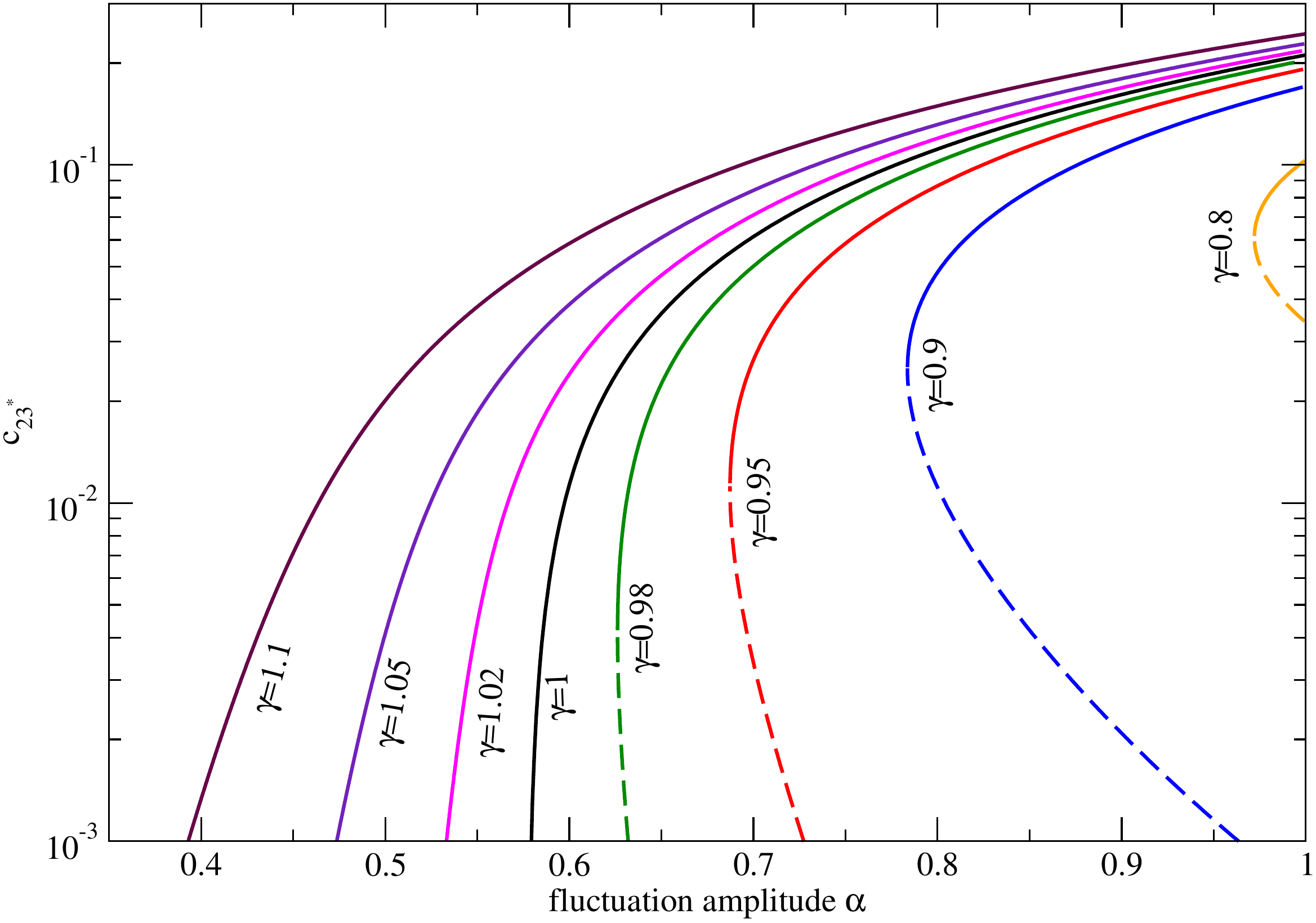}
\caption{\label{fig:triang_0}
Bifurcation diagram showing the fixed point values of conductance $C_{23}$ in the system with two sinks fluctuating at amplitude $\alpha$. An additional fixed point (not shown due to the logarithmic axis) is the trivial one with $C_{23}=0$ present for all values of $\alpha$. The three nodes form an equilateral triangle, $L_{12} = L_{13} = L_{23} = 1$. 
}
\end{figure}

\paragraph{Case $\gamma \neq 1$.}
For $\gamma \neq 1$ and with $C_{12} \neq 0$, we may solve \eqref{eq:stationary12} for $C_{23}$, obtaining
\begin{align}\label{eq:C23}
C_{23} = C_{12}  (4- C_{12}^{-\gamma-1})^{\frac{1}{\gamma-1}} L_{23}^\frac{1+\gamma}{1-\gamma}~.
\end{align}
Using \eqref{eq:XfromC23} and replacing $X$ according to \eqref{eq:pl_ph}, we obtain
\begin{align}\label{eq:alpha}
\alpha = (C_{12}+2C_{23}) C_{23}^\frac{\gamma-1}{2} L^\frac{\gamma+1}{2}.
\end{align}
Now \eqref{eq:C23} and \eqref{eq:alpha} implicitly provide stationary solutions for $\gamma=1$: insert a value of $C_{12}$ in \eqref{eq:C23} to obtain $C_{23}$; then insert
$(C_{12},C_{23})$ in \eqref{eq:alpha} to obtain the value of $\alpha$ for which $(C_{12},C_{23})$ is a stationary solution. Figure \ref{fig:triang_0} shows the resulting bifurcation diagram for various choices of $\gamma$. The curve for $\gamma=1$ is directly obtained from \eqref{eq:C23forgamma1finalequi}.

\begin{figure}
\includegraphics[width=\columnwidth]{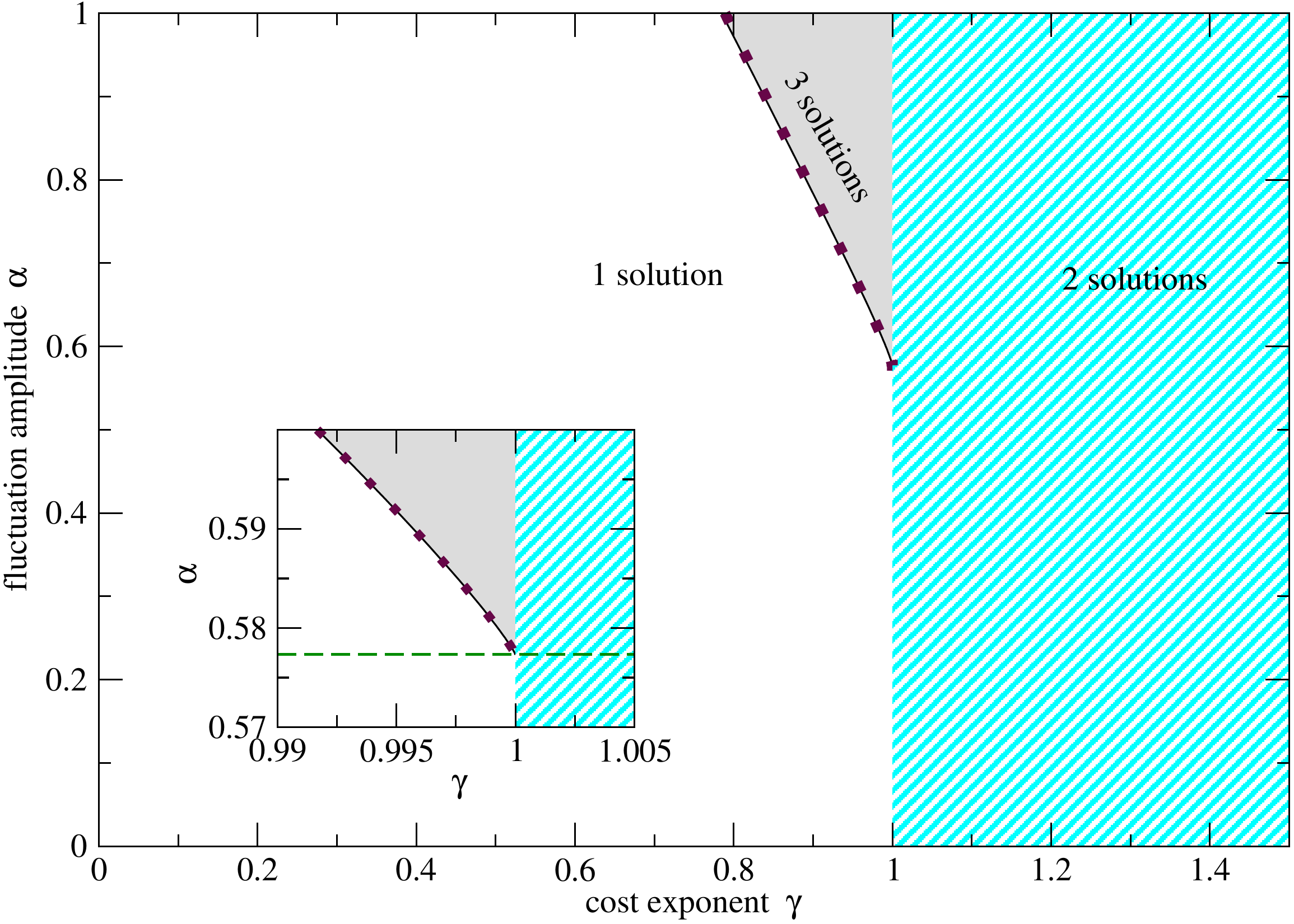}
\caption{\label{fig:trianglei_0}
Multiplicity of stationary solutions for the system with two sinks in the space of parameters $\gamma$ and $\alpha$ at $L_{12}=L_{23}=1$. Between parameter regions with 1 and with 3 solutions, a saddle-node bifurcation occurs, cf.\ curves with $\gamma<0$ in Figure~\ref{fig:triang_0}. The $(\alpha,\gamma)$ curve separating these two regions (thin dark curve) is approximated well (large dots) by the calculation leading up to \eqref{eq:C12_crit_approx}. The tree-like solution with $C_{23}=0$ always exists; it is stable for $\gamma<1$, unstable for $\gamma>1$; for $\gamma=1$ it changes stability in a transcritical bifurcation at $\alpha_c=\sqrt{1/3}$ (dashed vertical line in inset). 
}
\end{figure}

We approximate the critical parameter value $\alpha^\ast$ in dependence of $\gamma$. We insert \eqref{eq:C23} into \eqref{eq:alpha},
\begin{align}
    \begin{split}
\alpha   = &       \left[C_{12}^\frac{\gamma+1}{2} (4-C_{12}^{-\gamma-1})^\frac{1}{2} L_{23}^\frac{-\gamma-1}{2} \right. \\
         + & \left. 2 C_{12}^\frac{\gamma+1}{2} (4-C_{12}^{-\gamma-1})^\frac{\gamma+1}{2(\gamma-1)} L_{23}^\frac{(1+\gamma)^2}{2(1-\gamma)} \right] L_{23}^\frac{\gamma+1}{2}
         \end{split}
\end{align}
Simplifying and introducing $u:=C_{12}^{1+\gamma}$, we obtain
\begin{align}
\alpha & = \sqrt{(4u-1)} + 2 \sqrt{u} (4-u^{-1})^\frac{\gamma+1}{2(\gamma-1)} L_{23}^\frac{\gamma^2+\gamma}{\gamma-1}
\end{align}
The derivative of $\alpha$ with respect to $u$ reads
\begin{align}\label{eq:alphaderivative}
\begin{split}
\frac{\d \alpha}{\d u}   = & 2 (4u-1)^{-1/2} + L_{23}^\frac{\gamma^2+\gamma}{\gamma-1} \\
\times & \left[ u^{-1/2}(4-u^{-1})^\frac{\gamma+1}{2(\gamma-1)} + \frac{\gamma+1}{\gamma-1} u^{-3/2} (4-u^{-1})^\frac{-\gamma+3}{2(\gamma-1)} \right]
\end{split}
\end{align}
The first term in the square brackets in \eqref{eq:alphaderivative} becomes negligible in comparison to the second one when $\gamma \rightarrow 1$. We drop this first term in the following calculation staying accurate for values of $\gamma$ sufficiently close to 1. In this approximation, setting $\d \alpha / \d u=0$ and using $u\neq 0$ implies
\begin{align}\label{eq:derivativezero}
0  = & 2 + u^{-1} \frac{\gamma+1}{\gamma-1} (4-u^{-1})^\frac{1}{\gamma-1} L^\frac{\gamma^2+\gamma}{\gamma-1}~.
\end{align}
Since $C_{12} = \sqrt{1/3}$ in the stationary solution for $\gamma=1$, we replace the prefactor $u^{-1} = C_{12}^{-\gamma-1} \rightarrow 3$ by this asymptotic value to obtain 
\begin{align}
2  = & 3 \frac{1+\gamma}{1-\gamma} (4-u^{-1})^\frac{1}{\gamma-1} L^\frac{\gamma^2+\gamma}{\gamma-1}
\end{align}
and further
\begin{align}
\left(\frac{2(1-\gamma)}{3(1+\gamma)}\right)^{\gamma-1} = (4-u^{-1}) L^{\gamma^2+\gamma}
\end{align}
to arrive at
\begin{align}
u = \left(4-\left(\frac{2(1-\gamma)}{3(1+\gamma)}\right)^{\gamma-1} L^{-\gamma^2-\gamma}\right)^{-1}~.
\end{align}
We replace $u=C_{12}^{1+\gamma}$ and obtain
\begin{align} \label{eq:C12_crit_approx}
C_{12} = \left(4-\left(\frac{2(1-\gamma)}{3(1+\gamma)}\right)^{\gamma-1} L^{-\gamma^2-\gamma}\right)^{-1-\gamma}~.
\end{align}
for the approximated conductance $C_{12}$ at the bifurcation. Inserting this value of $C_{12}$ into \eqref{eq:C23} and \eqref{eq:alpha} provides conductance $C_{23}$ and the critical value $\alpha$ at the bifurcation in the regime of $\gamma$ close to 1. Figure~\ref{fig:trianglei_0} shows that the approximation works well even for values of $\gamma<0.9$.

\begin{figure}[htp!]
\includegraphics[width=\columnwidth]{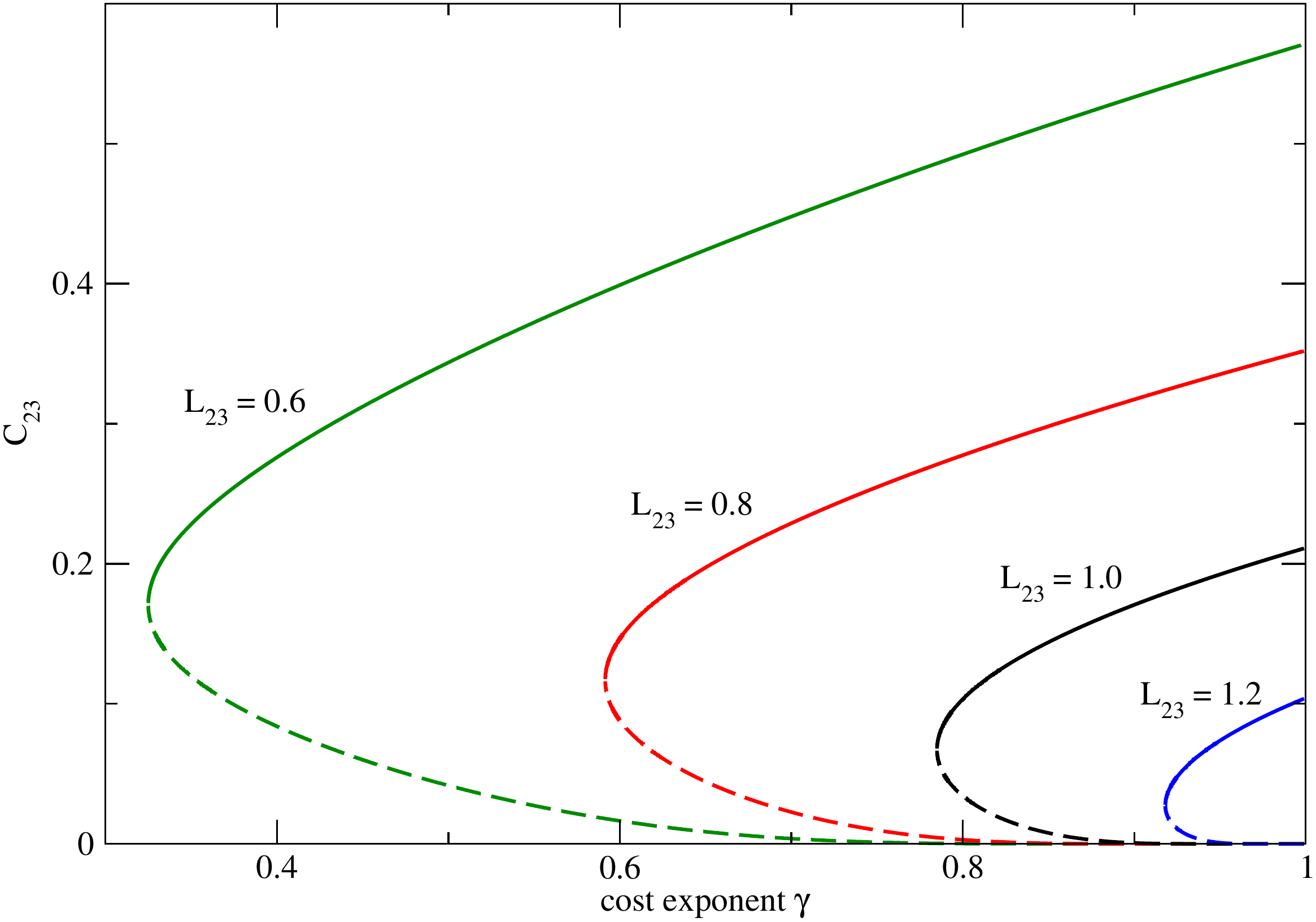}
\caption{\label{fig:triangleg_0}
Triangular stationary solutions under varying cost exponent $\gamma$ as a bifurcation parameter but constant maximal fluctuation amplitude $\alpha=1$. The diagram shows the fixed point values of conductance $C_{23}$ in the system with two sinks connected by a vessel segment of length $L_{23}$. The other two connections in the system have length $L_{12} = L_{13} = 1$. An additional fixed point is the trivial one with $C_{23}=0$ (not shown due to the logarithmic axis). 
}
\end{figure}

\paragraph{Bifurcations varying $\gamma$ for different length $L_{23}$.} We complete the consideration of the system with three nodes by observing response to varying $\gamma$ is varied while keeping $\alpha$ constant. This scenario, for different lengths of the cross-connection $L_{23}$, is the most relevant to the parameter scan with the empirical network in section \ref{sec:scanempirical}. Figure \eqref{fig:triangleg_0} shows a pair of a stable and an unstable triangular solution disappear in a saddle-node bifurcation as $\gamma$ decreases through a critical value. Solutions are obtained numerically from \eqref{eq:C23} and \eqref{eq:alpha} by tuning $C_{12}$ such that $\alpha=1$ results for $\gamma$ and $L_{23}$.

\subsection{Parameter scan on an empirical vascular network} \label{sec:scanempirical}

\begin{figure*}
 \includegraphics[width=\textwidth]{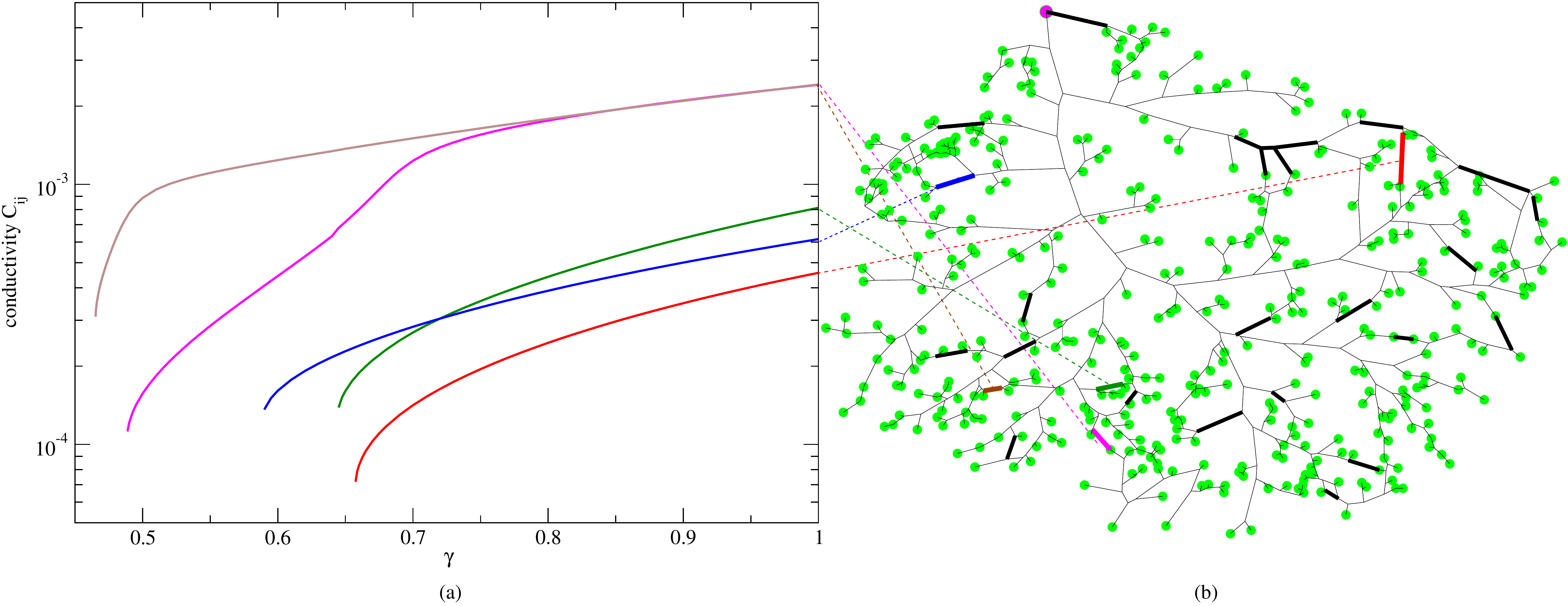}
\caption{\label{fig:empirical}
(a) Stationary solutions of conductances undergo bifurcations as the cost exponent $\gamma$ is varied in an empirical vascular network. The amplitude of fluctuations is kept constant at $\alpha=1$ (single moving sink). The 5 curves show the $\gamma$-dependence of  the five conductances vanishing earliest as the cost exponent is reduced from $\gamma=1$ towards $\gamma=0$. (b) In this drawing of the empirical vascular network, each node is placed according to its given coordinates. The source (root) is the topmost node marked with a filled circle. All other filled circles are sink nodes (having degree 1 or 2). A connection drawn by a thick line indicates that its conductance turns zero in a bifurcation as $\gamma$ decreases. A dashed line joins each of the five conductances featured in panel (a) with its connection drawn in panel (b).
Fixed points of the model equations \eqref{eq:adaptation} are found by Euler integration with a time step $\Delta t =10^{-3}$. Initial conductances are drawn from $[0,1]$ uniformly at random. 20 runs with independent initial conditions all produce the outcome displayed in panel (a). 
}
\end{figure*}

We analyze the stationary solutions of the model on the empirical vascular network described in section~\ref{sec:networkdata}. While keeping fluctuations at maximum ($\alpha=1$, single moving sink), we vary the cost exponent $\gamma$ and observe the conductances after settling to stable stationary solutions of \eqref{eq:adaptation}. 
We observe that for $\gamma=1$, all vessel segments have a positive conductivity, $C_{ij}>0$. As $\gamma$ decreases, there is a sequence of bifurcations in each of which a single connection turns from positive to zero conductance. Figure \ref{fig:empirical}(a) shows the $\gamma$-dependence of five conductances involved in bifurcation earliest, i.e.~ at relatively large $\gamma$.

As $\gamma$ decreases towards zero, a total of 30 bifurcations occur. In each of these, a single connection loses its non-zero conductance.  The 30 connections involved in bifurcations are drawn as thick lines in the network illustration in Figure~\ref{fig:empirical}(b). The 825 connections retaining non-zero conductance for $\gamma \rightarrow 0$ form a tree on the 826 nodes: by following these remaining connections, all nodes are mutually reachable by a unique path so there are not any cycles.

Empirically detected vessel segments naturally have a non-zero conductance. Assigning positive conductances to all existing vessel segments is required for a realistic parameter choice of the model. The result in Figure~\ref{fig:empirical} suggests that a choice of $\gamma=0.5$ \cite{HuCai2013} does not reproduce the empirical network with all positive conductances. However, choices of $\gamma$ closer to a value of $1$ do provide stable stationary solutions with all conductances positive.

Here we have assumed that the case of maximal fluctuation amplitude, $\alpha=1$, maximizes conductances in the stable stationary solutions. While this assumption is plausible in the light of the results obtained for the model so far, a demonstration of the assertion is left for future work.

Since empiricial data --- in general --- cannot be guaranteed to be free of error, one may ask if the conductances vanishing at large values of $\gamma$ belong to erroneously registered vessel segments. This appears unlikely when considering the length of the connections. The lengths of the five vessel segments featured in Figure~\ref{fig:empirical}(a) lie in the range $[28.6;79.6]$ to be compared to median $15.3$ and mean $20.7$ for the distribution of all 855 segments lengths in the network.

\section{Discussion}
% Summary.
The model for vascular network adaptation by Hu and Cai~\cite{HuCai2013} exhibits rich phenomenology in terms of bifurcations even in the simplest non-trivial system with one source and two fluctuating sinks. In the biologically most relevant regime of cost exponent $0 < \gamma < 1$, tree-like solutions are stable irrespective of other parameters. Cyclic (triangular) behaviour appears as a pair of a stable and an unstable solution in a saddle node-bifurcation. In the supercritical regime, there is multi-stability. Tree-like or cyclic behaviour of the system depends on initial conditions and basins of attraction in the space of conductances.

Testing the model on an empirical network with hundreds of nodes reveals cyclic solutions appearing by saddle-node bifurcations, just as in the 3-node system. In contrast to the rigorous analysis of small systems, we were able to find stable stationary solutions only by integrating the equations of motion from a random sample of initial conditions.  Asymptotics of the trajectories being consistent across initial conditions makes it plausible that we are identifying the stable fixed point whose basin of attraction is dominating the phase space.

In terms of model calibration, our test of the model on an empirical network suggests that the cost exponent is closer to a value of $\gamma =1$ than the previosuly suggested\cite{HuCai2013} value $\gamma = 1/2$. At the latter value, the model's prediction is incompatible with simultaneous presence of all empirically observed vessel segments. In the stable stationary solutions found at $\gamma=1/2$, a value of zero conductance is assigned to several vessel segments, incompatible with the existence of these connections. Increasing $\gamma$, however, the model assigns realistic non-zero conductances to all connections in agreement with them being observed as vessel segments in the empirical network.

Future work is to improve methods and aim at exhaustive identification of all stationary solutions also in large networks. Structural features of vascular networks are to be exploited by a suitable method. In particular, network sparseness may facilitate efficient search of solutions. Typical vascular networks have a small cyclomatic number, meaning all cycles can be removed by breaking a relatively small number of connections. The structural similarity to trees is also seen in low tree-width \cite{Bodlaender:2010} of vascular networks that we have identified in preliminary work \cite{KlemmMartens2023,KlemmJPC:2020}. Ideally, the tree-like structure can be exploited to recursively compute the set of stationary solutions on a network.

The robustness of results is to be tested by varying the type of sink fluctuations. While the implementation of fluctuations in the present paper is in line with our own previous work \cite{MartensKlemm2017,martens2019cyclic}, it differs from the original definition by Hu and Cai \cite{HuCai2013}. In the latter, each sink $i$ independently opens with a probability $p$, otherwise it fully closes. The case of maximal fluctuations ($p \rightarrow 0$), however, is that of a single moving (open) sink and thus identical to maximum fluctuation amplitude ($\alpha=1$) in the present work. Going beyond these two types of fluctuations, spatio-temporally correlated sinks and their effects on tree-like versus cyclic network structure are worth exploring.

% Applications.
Several situations are conceivable that prompt the need to use an model that adapts vessel diameters~\eqref{eq:adaptationrule}, towards finding stable  configurations (vessel diameters) that are consistent with flow parameters.
First, some experimental network data only provides information about topology (connectivity), but not about vessel diameters, including the data used in this study~\cite{Blinder2010} and elsewhere~\cite{lipowsky1974network}. However, in some circumstances such data may be relevant to add to simulations --- in such cases, the adaptive model~\eqref{eq:adaptationrule} can be used to generate data with vessel diameters that are consistent with prevalent flow parameters~\cite{mestre2020cerebrospinal}.
Large scale whole brain vascular networks measured in rodents have recently become available~\cite{kirst2020mapping}. While these data contain information about vessel widths, the data are extracted from animals {\it post mortem}. This raises two questions: first, does the data contain erroneous vessels (presumably small or short) --- this question relates to the consistence of the network data. Second, what are the actual {\it in vivo} diameters that appear when blood pressure corresponds to the alive state of the investigated organism. These questions are also be relevant when constructing realistic models of vasculature in whole organs~\cite{tithof2022network,Postnov2016}, and adaptive models such as the one studied here may be used to check for consistent edge weights (conductivities) along the network.

%  Outlook.
In future work, one may raise the question of relevance of certain vessels large networks where the size of the network is prohibitive for the computation/simulation of the entire Kirchhoff network; i.e., is it possible to simplify (i.e., coarse grain) the network towards computational accessibility? Moreover, myogenic responses are known to lead to oscillatory behavior in vessel diameters. Are such oscillations a feature of the self-organization intrinsic to the adaptive dynamic nature of the network? Or are these oscillations due to external forcings due to neuronal activity (astrocytes and pericytes)?
Finally, it would also be interesting to conduct experiments with flow networks. An example of a potential candidate system is an experimental system with adaptive memristor flow circuits proposed very recently~\cite{martinez2023fluidic}.
 
% % 
% We conclude that the model can capture some aspects of vascular network formation and adaptation, but also suggest some limitations and directions for future research. Our findings contribute to a general understanding of the dynamics in adaptive transport networks, which is essential for studying mammalian vasculature and developing self-organizing piping systems.

\begin{acknowledgments}
K.K. acknowledges support from Project No. PID2021-122256NB-C22 funded by
MCIN/AEI/10.13039/501100011033/FEDER, UE.
\end{acknowledgments}

\section*{Data Availability Statement}

Data sharing is not applicable to this article as no new data were created or analyzed in this study.

\appendix

\section{Appendixes}
\subsection{'Power' terms}\label{app:fluctuations} It is useful to determine expressions for the forcing terms appearing in the triangular network system~\eqref{eq:triangular_motif_goveqn_symmetric_b}, $ \langle h_2^2 \rangle $, $\langle (h_2-h_3)^2 \rangle $ and $\langle h_2h_3\rangle$,  as functions of the fluctuation amplitude, $a$.  To this end, we consider two types of forcing.

\paragraph{Periodic drive.} Suppose $h_{2,3} = -1/2 \pm a/\sqrt{2}\cos{(\omega t)}$. Averaging with $\langle f \rangle_T = T^{-1}\int_0^T f(t)\d t$, we then have
\begin{subequations}
 \begin{align}\label{eq:fluctuations_periodic}
  \langle h_2^2 \rangle &=  \frac{1+a^2}{4},\\
  \langle (h_2-h_3)^2 \rangle &=  a^2,\\
    \langle h_2 h_3 \rangle &=  \frac{1-a^2}{4}.\
\end{align}
\end{subequations}
%  ----------
\paragraph{Stochastic drive.} We define a high and low flow, $h^+= -\half(\alpha+1)$, $h^-= \half(\alpha-1)$), that apply to either node 2 or 3, respectively.
\begin{subequations}
\begin{align}\label{eq:fluctuations_stochastic}
\nonumber
    \langle h_2^2 \rangle &= 
    p_l\cdot (h^+)^2+
    p_r \cdot (h^-)^2\\
%     &= 
%     \frac{1}{2} ((h^+)^2+ (h^-)^2)\\
%     &=
%     \frac{1}{2} \left(\left(\frac{\alpha+1}{2}\right)^2+ \left(\frac{\alpha-1}{2}\right)^2\right)\\
    &=\frac{1}{4}(1+\alpha^2)
    \\
% %     
\nonumber
    \langle (h_2-h_3)^2 \rangle &= 
    p_l \cdot (h^+ - h^-)^2 + p_r \cdot (h^- - h^+)^2, \\
%     &= \frac{1}{2}(\alpha^2 + \alpha^2) \\
    &= \alpha^2, \\
% %     
\nonumber
    \langle h_2 h_3 \rangle 
    &=  p_l \cdot (h_+h_-) + p_r \cdot (h_-h_+) \\
%     &= \frac{2}{2} h_+h_-\\
    &=\frac{1}{4}(1-\alpha^2).
\end{align}
\end{subequations}

% \bibliography{artdata}

\end{document}